# Advanced Persistent Threat: Detection and Defence

Mohammad Bilal Khan
Faculty of Engineering & Informatics
University of Bradford
Bradford, UK
M.B.Khan6@bradford.ac.uk

*Abstract*—The critical assessment presented within this paper explores existing research pertaining to the Advanced Persistent Threat (APT) branch of cyber security, applying the knowledge extracted from this research to discuss, evaluate and opinionate upon the areas of discussion as well as involving personal experiences and knowledge within this field. The synthesis of current literature delves into detection capabilities and techniques as well as defensive solutions for organisations with respect to APTs. Higher-tier detection and defensive strategies bear greater importance with larger organisations; especially government departments or organisations whose work impacts the public on a large scale. Successful APT attacks can result in the exfiltration of sensitive data, network down time and the infection of machines which allow for remote access from Command-and-control (C2) servers. This paper presents a well-rounded analysis of the Advanced Persistent Threat problem and provides well-reasoned conclusions of how to mitigate the security risk.

## I. INTRODUCTION

As defined by the National Cyber Security Centre of the United Kingdom, an Advanced Persistent Threat or APT is a "targeted cyber-attack where a hacker accesses a system and remains undetected for a long time" [3]. From this definition alone, the high likeliness of malicious actors using a variety of infiltration techniques to compromise targets can be deduced, many of which will be mapped out within the MITRE ATT&CK Framework. With the advancement of modern-day detection techniques, it can be understood that the complexity and sophistication of code produced for these attacks to remain undetected must be high. As the APT are well funded, organized groups, the production of flawless code for delivering payloads come as no issue to these groups [9].

To understand defensive solutions, a good understanding of the motivation empowering APT groups and their attacks is required. Within the IEEE Security & Privacy journal, it is described how frequently, the primary objective behind these attacks is money and the secondary objectives include national and industrial espionage [2, 21]. A very famous example of national espionage is that of APT28 whereby, spear-phishing attachments were used to compromise the email accounts of government officials during the 2016 United States Election. The compromise of Hilary Clinton's campaign following leaked emails are said to have impacted the results within the election [11].

Industry-recognized security expert, Dr, Eric Cole explains how organisations understand the traditional defensive strategies when such as firewalls. However, these traditional approaches experience no success when met with an APT attack due to these attacks being a completely different problem [4, 9]. From this, it can be argued that organisations are struggling to keep up with changes within the cyber-industry with respect to the nature of attacks that are present. Greater education and emphasis on emerging vulnerabilities are things organisations should bear in mind. Implementing detective and defensive solutions to specifically deal with APT attacks as well as ensuring staff working in cyber analyst positions understand what an APT is and how the organisations defensive strategies are structured in the event they are targeted by an APT group [6].

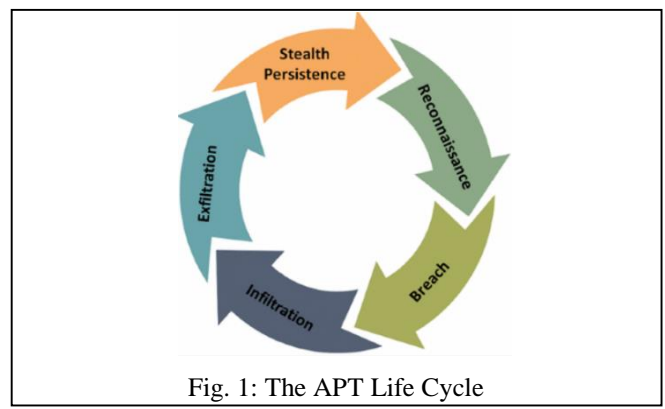

Fig. 1: The APT Life Cycle

Saurabh Singh [19] describes the five crucial stages to the life cycle of an APT attack. The first step being reconnaissance. This is where research pertaining to the organisation and its employees. Social networking sites such as LinkedIn help to provide the attackers with the information they require. The second step is breach. This is where we see malware, SQL injections, zero-day vulnerabilities as well as other methodologies being used to gain a foothold into the network and breach defence systems [8]. The third step is infiltration. Now we see the attacker searching for confidential data and company documents. They will move horizontally within the network to identify servers and user data. This is done with the help of powerful software tools. Moving to step 4, exfiltration will take place. The APT group will take control of the network and thus, leak confidential company data, including intellectual property rights. Data will be transferred from the company's network to an external location, most likely a Command-and-Control centre in which the actor(s) will control. In the final stage of an APT attack, we tend to see stealth persistence being achieved. Providing that the exfiltration of data was silent, attackers many choose to remain inside the network which will allow them to obtain further confidential documents from the organisation over a period [10].

Throughout the discussion, references to the MITRE ATT&CK Framework have been made and will continue to be made. This framework breaks down the stages of an attack, from initial access up till exfiltration and impact. During the discussion of detective and defensive solutions, references to the corresponding part of the framework will be made to show which stage of an attack that solution tackles. The MITRE Corporation describes the framework as a foundation for developing specific threat models and methodologies within the cybersecurity product and service community [11].

## II. DETECTION/DEFENSE TECHNIQUE – ANOMALY BASED INTRUSION DETECTION SYSTEMS

### A. OUTLINING ANOMALIES

What needs to be considered when discussing APT attacks is that these attacks are unique and tailored. Traditional security methods are bypassed as a result. Anomaly detection can be defined as the problem of discovering patterns within data that do not conform to what would normally be expected [5]. Thus, the ability for anomaly detection within intrusion detection systems would contribute to greater success in scalping APT attacks. It would provide the opportunity to analyse irregular behaviour in the form of abnormal patterns within data. Ivo Freidberg [12] describes how anomalies can be categorised into three categories.

1) Point anomalies whereby a single event is identified as anomalous with respect to how normality is viewed.

2) Contextual anomalies whereby a single event is identified as anomalous with respect to the context it was identified within.

3) Collective anomalies whereby the relation formed collectively between a series of events places them within an anomalous bracket.

### B. DETECTING ANOMALIES

Now that a base understanding for anomalies has been adopted, the discussion of how to detect these anomalies can be brought up. There are various anomaly detection algorithms that exist, many of which are predominantly known and many of which have been mentioned in journals and articles, developed by specialists within the computational geometry and cyber-security field.

An example of one of these algorithms is Classification based anomaly detection. This technique uses a classifier or model to study a labelled training data set. The learned model then classifies a test instance into a class, that being either normal or anomalous through using the classifier [5]. The general assumption for this algorithm describes the ability for a classifier to distinguish between normal and anomalous instances of data [12].

### C. VIABILITY OF AN ANOMALY BASED INTRUSION DETECTION SYSTEM FOR APT ATTACKS

A controlled test conducted by Min Qin and Kai Hwang [15] from the University of Southern California using an anomaly-based intrusion detection system showed that the system was able to find unknown attacks embedded within telnet, http, ftp and smtp as well as other TCP, UDP or ICMP connections. A detection success rate of 47% for denial of service attacks, remote-to-local and probe attacks was witnessed. In the words of Min and Qin themselves, these results prove the viability of an anomaly-based IDS.

The major drawbacks which comes with the implementation of an anomaly-based IDS is the number of false-positives. Any data instance that doesn't perfectly align with the normal class gets categorised as anomalous. Within a cyber-security working environment whereby workload is often very busy, having to deal with a large instance of false positives daily wastes valuable time. As somebody who has worked in cyber security for the government for a year as part of an industrial placement, I myself have witnessed the number of false positives from different sensors that have to be dealt with. Introducing an anomaly-based IDS which will add on top of the current situation, even more false-positives will create a heavier workload for cyber analysts. However, looking at this, it could be argued that this could help to create and enforce more jobs within firms [13, 14].

Timothy Shimeall [17] argues that the lack of granularity is an issue. The system picks up what it believes to be anomalous however, it is not for definite. This ties into the point made in the paragraph above which is the high production of false-positives. Furthermore, we can tie in how the success of the intrusion detection system is heavily dependent upon the surroundings in which it learns to categorise what normal is. If this stage is rushed or not done correctly, false-positive rates could be even higher.

To conclude on anomaly-based intrusion detection systems, they are clever systems that implement computational geometry methods and techniques to identify outliers within data instances and report this back to the cyber team of a company [16]. As APT attacks are unique and tailored to companies, the implementation of this system will definitely help to detect these attacks and at least make cyber teams aware of what is happening, which could help them to fight against the stealth persistence stage of the APT cycle in the worst-case scenario that confidential documents are exfiltrated. Looking at the MITRE ATT&CK Framework, the success of an anomaly-based IDS will aid companies at the Initial Access stage of the framework, meaning the first stage. The Initial Access stage can be said to be equivalent to the breach stage of the APT life cycle. This system would be very valuable to companies for this very reason. The downside to this system is the high false alarm rate [18]. Time is so valuable to cyber analysts and increasing their workload to deal with a handful of false positives daily may not be the best idea. It completely depends upon the nature of the company, the size, the financial capabilities and staffing capabilities.

## III. DETECTION/DEFENCE TECHNIQUE – USER AWARENESS OF SPEAR-PHISHING EMAILS

### A. OUTLINING SPEAR-PHISHING

APT groups have learned that servers are harder to break into due to being locked down. What is more effective is to target the users who have access to this information [9]. It is not hard for malicious actors to obtain personal information about several high-level employees within a firm. This is due to the amount of information we reveal about ourselves on social media, specifically LinkedIn and Facebook. As mentioned earlier, I worked for the British government within cyber security. Many reconnaissance emails were witnessed daily whereby users who had revealed personal information about themselves on LinkedIn were being targeted with blank emails. You might ask, why would a malicious actor send a blank email? Sometimes, it is hard for actors to obtain the exact email address whereby you receive your business emails [20]. However, so long as they know your name and the company you work for, they can discover the email domain and predict the start of your email address. They then send reconnaissance emails which is a blank email from a non-suspicious looking address. Many users, being oblivious to spear-phishing, reply to these emails asking whether this email was meant for them. Once the malicious actor receives a reply, they know they have an active address for their target.

Symantec, a well-known cyber security software company reported that spear-phishing attacks in the form of business emails are being favoured by malicious actors. Looking at this, it can be deduced that one of the underlying causes will be the simple fact that humans are easier to compromise than security systems. Through social engineering, even the most security savvy employee can be fooled [1].

### B. CONTROLLING OUR PERSONAL INFORMATION

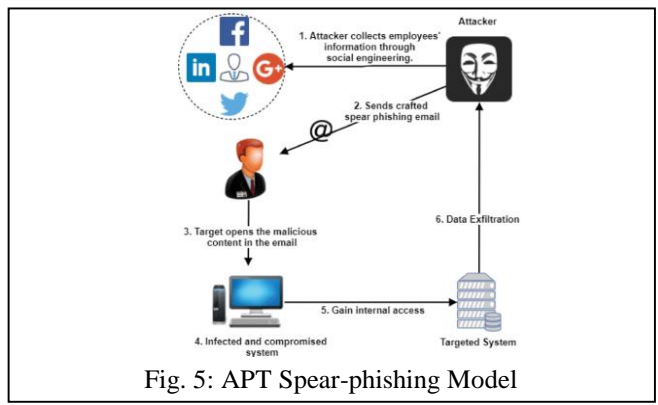

Fig. 5: APT Spear-phishing Model

Figure 5 above demonstrates the APT life cycle within the context of a spear-phishing attack. What needs to be focused on is stages one and two. By diminishing the ability for attackers to collate any background information on employees, the company becomes a harder challenge to malicious actors who may even reconsider targeting that company. This can be done through a variety of techniques. As somebody who has worked for the government, the following advice is coming from personal experience working in a highly controversial organisation. The best strategy would be to enforce strict company policies. The type of policies which bear disciplinary actions. The first policy should be the non-usage of company email addresses for personal usage. Using a company email address to sign up to non-work-related sites should be banned. Within today's day and age, everybody should have access to a personal address they can use for their own activities online. Work matters and personal matters should not mix [22, 23]. The second policy that should be enforced really depends upon the type of company you work for. Although you are aware that I worked for the British government, you are not aware which exact government department I was a part of. High profile companies who have great political impact or who make decisions that may cause controversy should make it obligatory on employees to not disclose their exact employment details online such as on Facebook and LinkedIn. However, more general employment details can be disclosed such as their experience within the manufacturing industry.

### C. EDUCATING EMPLOYEES ON THE ISSUE

Unfortunately, spear-phishing attacks still take the prize for being the most successful attack vector on enterprises [24]. Dr Eric Cole [5] explains how the number one reason for organisations becoming compromised today is lack of user awareness. Due to this, they open attachments that they should not. APT groups take advantage of this.

Looking at this, it can be seen how more focus needs to be placed on making users aware of what spear-phishing is. When was the last time you worked for a company and was given advice on dealing with spear-phishing? Never, right? If users are unaware of the issue, how can it be expected of them to deal with it in the correct manner.

As part of the industrial placement within the government, I took part in Learning at Work days whereby I taught and presented to new employees the issue of spear-phishing. This is a key point. The interaction between cyber analysts and other teams needs to be established. Within organisations, what tends to happen is you stay within the bubble which is your team and you don't really get to learn about what other teams do and the importance of their work. More emphasis needs to be placed on the work cyber analysts do. It should be well-known to all staff the hard work cyber-analysts commit to in order to protect the organisation. One of the ways this can be done is through creating that connection between the cyber team and other teams within the organisation. The ability to pass on critical information such as being mindful of opening attachments and enabling macros within spreadsheet files is what could save an organisation from being compromised by an APT attack.

The second thing that was taken away from the placement which could benefit other organisations is sending out test phishing emails. Think about it. You can describe a phishing email and all the signs to look out for as best as you can, however, until an employee sees a real phishing email, this will decipher whether user-education and awareness is spreading throughout the organisation. Several of these test emails would be sent to users throughout the year, mimicking hotel booking companies, florists, clothing retailers. If the user clicked on the URL, they would be taken to a site informing them about what they have just done and how the URL they clicked on could have been malicious. This is usually enough to make employees vigilant about what they click on and open within emails. If the user forwarded the email to the phishing team within the cyber security department, they would receive a thank you email explaining to them how their vigilance helps to make our organisation a safer environment.

Looking at the MITRE ATT&CK Framework, the success of strict policies and user education will aid companies at the Initial Access stage of the framework. Looking at the APT life cycle, implementing these defence strategies successfully will protect against the reconnaissance and breach stages.

### IV. CONCLSION

Through the research and personal experience working with these defence strategies, both techniques bear significant and critical importance. The anomaly-based IDS gives the company the ability to pick up where unusual activities occur within their infrastructure. Obtaining the IDS logs within the SIEM tool, analysts can reduce data instances to more readable formats as well as pin-point monitoring on certain data instances. It can give companies that confidence knowing they have a system in place for picking up behaviour that doesn't correlate to what they would expect. This technique focus purely on network monitoring, placing IDS sensors on servers, outlining normal behaviour and creating classes based on this knowledge. The strategy is very technical.

On the contrary, the implementation of strict company policies and user-education focuses solely on human psychology. The ability to give the workforce what they need to not fall victim to social engineering. This strategy is not technical but rather, it deals with the issue of APT attacks being able to take advantage of the blind-trust humans have of emails with embedded URLs and attachments.

Both defence strategies take two separate routes to tackle the same problem, which referring to the MITRE ATT&CK Framework, is the Initial Access stage. Being able to prevent APT groups gaining a foothold into the network infrastructure and move laterally. As mentioned earlier, both are critical for preventing successful APT attacks. To choose between financial investment between the two would be a hard task for any security manager. However, as Dr Eric Cole [9] said, APT attacks have become the primary focus for organisations but the traditional threat remains. Focus should not be placed on protecting against the APT attack to the point where other vulnerabilities and indicators of compromise are neglected.